\documentclass{www2010-submission}

\newfont{\mycrnotice}{ptmr8t at 7pt}
\newfont{\myconfname}{ptmri8t at 7pt}
\let\crnotice\mycrnotice%
\let\confname\myconfname%

\usepackage{multirow}
\usepackage{leqno}
\usepackage{graphicx}
\usepackage{url}
\usepackage{subfigure}
\usepackage{amssymb}
\usepackage{float}
\usepackage{marvosym}
\usepackage{epstopdf}
\usepackage{epsfig}
\usepackage{array}
\usepackage[pass,paperwidth=8.5in,paperheight=11in]{geometry}
\begin{document}


\title{Access Patterns for Robots and Humans in Web Archives}
\numberofauthors{1}%
\author{
\alignauthor
Yasmin AlNoamany,
Michele C. Weigle,
Michael L. Nelson\\
 \affaddr{Old Dominion University}\\
 \affaddr{Norfolk, VA, USA}
 \email{\{yasmin, mweigle, mln\}@cs.odu.edu}
}
\maketitle

\newcommand{\squishlist}{
  \begin{list}{$\bullet$}
  {
   \setlength{\itemsep}{0.9pt}
   \setlength{\parsep}{0.5pt}
   \setlength{\topsep}{0.5pt}
   \setlength{\partopsep}{0pt}
   \setlength{\leftmargin}{1.5em}
   \setlength{\labelwidth}{1.5em}
   \setlength{\labelsep}{0.5em} 
   } 
 }

\newcommand{\squishlisttwo}{
  \begin{list}{$\bullet$}
  {
   \setlength{\itemsep}{0.9pt}
   \setlength{\parsep}{0.1pt}
   \setlength{\topsep}{0.1pt}
   \setlength{\partopsep}{0pt}
   \setlength{\leftmargin}{1.5em}
   \setlength{\labelwidth}{1.5em}
   \setlength{\labelsep}{0.5em} } }

\newcommand{\squishend}{
   \end{list}  }
\begin{abstract}
Although user access patterns on the live web are well-understood, there has been no corresponding study of how users, both humans and robots, access web archives. Based on samples from the Internet Archive's public Wayback Machine, we propose a set of basic usage patterns: Dip (a single access), Slide (the same page at different archive times), Dive (different pages at approximately the same archive time), and Skim (lists of what pages are archived, i.e., TimeMaps). Robots are limited almost exclusively to Dips and Skims, but human accesses are more varied between all four types. 
Robots outnumber humans 10:1 in terms of sessions, 5:4 in terms of raw HTTP accesses, and 4:1 in terms of megabytes transferred. Robots almost always access TimeMaps (95\% of accesses), but humans predominately access the archived web pages themselves (82\% of accesses). In terms of unique archived web pages, there is no overall preference for a particular time, but the recent past (within the last year) shows significant repeat accesses.
\end{abstract}



\section{Introduction}
The web has become an integral part of our lives, shaping how we get news, shop, and communicate. In turn, web archives have become a significant repository of our recent history and cultural heritage. The Internet Archive \cite{wayback:billion} is the largest and oldest of the various web archives, holding over 240 billion web pages with archives as far back as 1996 \cite{Kahle2013}. Access to this vast archive is available through the Wayback Machine \cite{Tofel2007}, which sees about 82 million requests per day, based on our dataset.

Previous work has studied how users access the live web \cite{Tsagkias:2012:LIM:2348283.2348330} and search engines \cite{Jansen:2006:WSW:1138797.1138813}, but few studies have investigated how users access web archives. Understanding the current demand for access to web archives can provide insights into how to make the best use of limited archiving and access resources.

In this paper, we provide an analysis of user accesses to a large web archive. We examine a set of anonymized Wayback Machine server access logs from February 2012. We investigate the differences between human and robot accesses of the Wayback Machine, identify four major web archive access patterns, and uncover the temporal preference for web archive access. In particular, we find that robots (such as crawlers and spiders) account for 91\% of all sessions and 93\% of all page requests. Yet, robots only outnumber humans 5:4 in terms of raw, unfiltered requests and 4:1 in terms of megabytes transferred. Humans download more information per session, as they typically download embedded resources (e.g., images and stylesheets), which robots ignore.

We introduce four basic user access patterns of web archives: Dip, Slide, Dive, and Skim. Dip is requesting for a single URI. Slide is browsing different archived copies of the same URI. Dive is following the hyperlinks of a page, but staying near the same datetime. Skim is requesting only the TimeMaps (list of all archived copies for a specific original resource). Robots exhibit the Dip and Skim patterns equally, both about 49\% of their sessions, and almost exclusively request TimeMaps. Humans exhibit the Dip (39\%) and Dive (30\%) patterns the most and access archived pages significantly more than TimeMaps.

This paper is organized as follows. Definitions of important terms and a review of related work on web usage mining and web archive studies are presented in Section 2. Section 3 contains the patterns for accessing web archives along with an explanation and example for each pattern. A description of the Wayback Machine's web server logs, the dataset we used in the analysis, and the methodology of this study are presented in Section 4. Section 5 contains the results from analyzing the Wayback Machine access logs. Future work and conclusions are presented in Section 6.

\section{Related Work}
Despite the significance of web archives in preserving web heritage, the aspect of web archive usage has been overlooked. The only previous related work is a study of the search behavior characterization for web archives \cite{Costa2011}. We highlight this work, but first we define the terms for our discussion. 

\subsection{Memento Terminology}
Memento \cite{nelson:memento:tr} is an HTTP protocol extension which enables time travel on the web by interlinking the current resources with their prior state. Memento introduces content negotiation in the datetime dimension using a special HTTP header, Accept-Datetime \cite{memento:rfc}. Memento defines the following terms:
\begin{itemize}
 \item URI-R denotes the original resource. It is the resource as it used to appear on the live web; it may have 0 or more mementos (URI-Ms).
\item URI-M is an archived snapshot for the URI-R at a specific datetime, which is called Memento-Datetime. e.g., URI-M$_{i}$= URI-R$\MVAt t_{i}$.
\item URI-T denotes a TimeMap, a resource that provides a list of mementos (URI-Ms) for a URI-R with their Memento-Datetimes, e.g., $URI-T(URI-R) = \{URI-M_{1}, URI-M_{2}, ..., URI-M_{n}\}$. (We will also refer to a TimeMap as TM URI-R, to emphasize the URI-R in later examples)
 \end{itemize}
Although we use Memento terminology, the logs we analyze are from the public access Wayback Machine and not the Memento API.
\subsection{Web Usage Mining}
The breadth and depth of research in the area of web usage mining is massive and increasing \cite{Aye2011, Kumar2010, Sisodia2012, Catledge1995}. Web usage mining involves discovering usage patterns from web data using data mining \cite{Srivastava2000}. The results obtained from web usage mining can be used in different applications, such as web traffic analysis, site modification, system improvement, personalization, business intelligence, and usage characterization. Our study provides traffic analysis and usage characterization by providing abstract models for accessing web archives.
 
Adams et al.\@ explored the usage patterns of scientific and historical data repositories \cite{adams-ssrctr-11-01}. However, their study focused on a variety of archive types (e.g., public vs.\@ private, digital but non-web resources) and does not directly address the issue of archiving the web. The only web usage mining research that has been conducted on the usage of web archives is the study of search behavior characterization of web archives based on a quantitative analysis of the Portuguese Web Archive (PWA) search logs \cite{Costa2011}. The authors introduced a comparison between search patterns of web archives and web search engines. Despite the different information needs for web archives and web search engine users, the search patterns for web archives had shown adoption of web search engine technologies. They found that most web archive users conducted short sessions. In our study, the sessions that are composed of one request contribute the most to the number of sessions. One important finding from analyzing the search interactions of the PWA logs is that the users prefer older documents. This is in contrast to what we found, that web archive users have significant repetitions for requests in 2011 (the year prior to our sample).

The challenge that faces web usage mining is detecting the robots who camouflage their identity and pretend to be humans. The robot detection problem has been examined in several studies \cite{Tan2002, Dikaiakos2005, Kwon2012, Guo2012}. Doran et al.\@ characterized robot detection techniques into four categories: syntactical log analysis, traffic pattern analysis, analytical learning techniques, and Turing test systems \cite{Doran2010}. We used syntactical log analysis (simple processing by finding the self-identified robots) and traffic pattern analysis (specifying features for contrasting robots with humans).
\begin{figure*}[p]
\centering 
	\subfigure[URI-T]{
	\includegraphics[scale = 0.5]{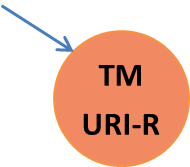}
	\label{fig:P01}
	}
	\subfigure[URI-M]{
	\includegraphics[scale = 0.5]{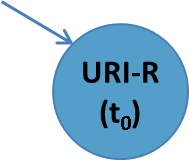}
	\label{fig:P02}
	}
\begin{tabular}{ |l|l|l| }
  \hline
  0.100.61.20 & 02/Feb/2012:06:48:24 & http://wayback.archive.org/web/*/http://iyasizuku.com \\ \hline  \hline 
  0.1.134.90 & 02/Feb/2012:07:08:28 & http://web.archive.org/web/19961022174810/http://altavista.com \\ \hline
\end{tabular}
\caption{Dip: A simple access to either a TimeMap or a memento.}
	\label{fig:P0}
\end{figure*}
\begin{figure*}
\centering
	\includegraphics[scale = 0.5]{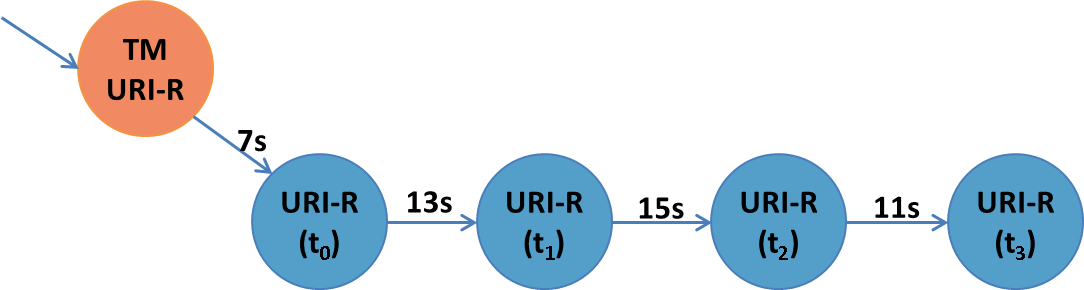}
\begin{tabular}{ |l|l|l| }
  \hline
  0.248.211.54 & 02/Feb/2012:07:04:52 & http://wayback.archive.org/web/20000715000000*/http://google.com  \\ \hline
  0.248.211.54 & 02/Feb/2012:07:04:59 & http://web.archive.org/web/20000301105534/http://google.com/  \\ \hline
  0.248.211.54 & 02/Feb/2012:07:05:12 & http://web.archive.org/web/20051101145803/http://www.google.com  \\ \hline
  0.248.211.54 & 02/Feb/2012:07:05:27 & http://web.archive.org/web/20080730200402/http://www.google.com/  \\ \hline
  0.248.211.54 & 02/Feb/2012:07:05:38 & http://web.archive.org/web/20110215024256/http://www.google.com/  \\ \hline
\end{tabular}
	\caption{Slide: Accessing the same URI-R at different Memento-Datetimes.}
	\label{fig:slide}
\end{figure*}
\begin{figure*}
\centering
	\includegraphics[scale = 0.5]{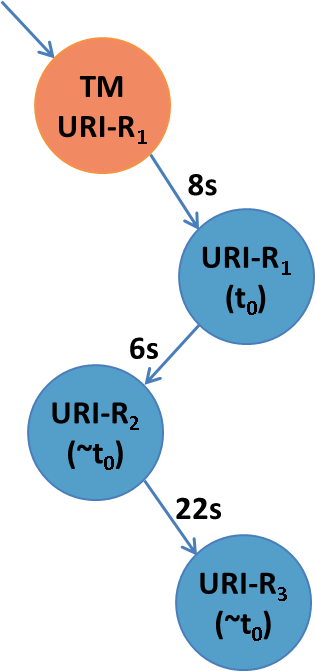}
	\begin{tabular}{ |l|l|l| }
  \hline
  0.106.160.155 & 02/Feb/2012:07:07:10 & http://wayback.archive.org/web/*/http://my-ru.net  \\ \hline
  0.106.160.155 & 02/Feb/2012:07:07:18 & http://web.archive.org/web/20100709124643/http://my-ru.net/  \\ \hline
  0.106.160.155 & 02/Feb/2012:07:07:24 & http://web.archive.org/web/20100709124643/http://my-ru.net/home.php  \\ \hline
  0.106.160.155 & 02/Feb/2012:07:07:46 & http://web.archive.org/web/20100706170736/http://my-ru.net/carousel.php  \\ \hline
\end{tabular}
	\caption{Dive: Browsing different URI-Rs at (approximately) the same Memento-Datetime.}
	\label{fig:P2}
\end{figure*}
\begin{figure*}
	\centering
	\includegraphics[width = 2in]{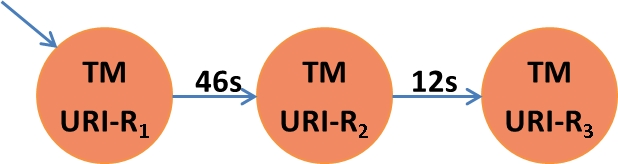}
\begin{tabular}{ |l|l|l| }
  \hline
0.10.212.177 & 02/Feb/2012:06:45:24 & http://wayback.archive.org/web/*/laquadrature.net  \\ \hline
0.10.212.177 & 02/Feb/2012:06:46:10 & http://wayback.archive.org/web/*/parti-du-plaisir.com  \\ \hline
0.10.212.177 & 02/Feb/2012:06:46:22 & http://wayback.archive.org/web/*/humanite.fr  \\ \hline\end{tabular}
	\caption{Skim: Traversing only TimeMaps for different URI-Rs.}
	\label{fig:P4}
\end{figure*}
\begin{figure*}
\begin{verbatim}
0.247.222.86 - - [02/Feb/2012:07:03:46 +0000] "GET http://wayback.archive.org/web/*/http://www.aura.vu 
HTTP/1.1" 200 96433 "http://www.archive.org/web/web.php" "Mozilla/5.0 (Macintosh; Intel Mac OS X 10_6_8)
AppleWebKit/535.7 (KHTML, like Gecko) Chrome/16.0.912.77 Safari/535.7" 

0.247.222.86 - - [02/Feb/2012:07:03:55 +0000] 
"GET http://web.archive.org/web/20020404020224/http://www.aura.vu/ HTTP/1.1" 200 18875 
"http://wayback.archive.org/web/*/http://www.aura.vu" "Mozilla/5.0 (Macintosh; Intel Mac OS X 10_6_8)
AppleWebKit/535.7 (KHTML, like Gecko) Chrome/16.0.912.77 Safari/535.7"}
\end{verbatim}
	\caption{Sample of the Wayback Machine access log.}
	\label{fig:sample}
\end{figure*}

\section{Abstract Models for Accessing Web Archives}
Through our analysis, we discovered four major patterns for web archive access. We present the model for each pattern along with an example from the logs in Figures \ref{fig:P0}-\ref{fig:P4}. Each example consists of three columns: the client IP, the access time, and the requested URI. The times annotating the transition arrows in Figure \ref{fig:slide}-\ref{fig:P4} represent the inter-request time in the given examples. Note that we use TM URI-R to denote a TimeMap in the figures. We use Memento terminology (URI-T, URI-M, and URI-R) in the definitions. We refer to the original resource for URI-T and URI-M with URI-R(URI-T) and URI-R(URI-M), respectively.
\subsection{Pattern 1: Dip}
Dip is the pattern where a user accesses only one URI. The request can be for a URI-T (Figure \ref{fig:P01} and the first example) or a URI-M (Figure \ref{fig:P02} and the second example). 
\begin{eqnarray}
Dip = \{\mbox{URI-X}_i|\:i=1\mbox{ and URI-X} \in \mbox{\{URI-T, URI-M}\}\} \nonumber
\end{eqnarray}

\subsection{Pattern 2: Slide} 
Slide is the pattern in which a user accesses the same URI-R at different Memento-Datetimes. In this pattern, the user requests a URI-R and walks through time browsing its different copies (Figure \ref{fig:slide}). 
\begin{eqnarray}
Slide &=&\{\mbox{URI-X}_i|\:i>1\mbox{, URI-X}\in \mbox{\{URI-T, URI-M}\} \nonumber \\
	  & & \mbox{and URI-R(URI-X}_{i}) = \mbox{URI-R(URI-X}_{i-1})\} \nonumber
\end{eqnarray}
Navigation between different URI-Ms can be done in many ways, e.g., directly from URI-M$_{1}$ to URI-M$_{2}$ (URI-R$\MVAt t_{1} \Rightarrow$ URI-R$\MVAt t_{2}$) or from URI-M$_{1}$ to URI-M$_{2}$, but in the middle the user returns to the TM URI-R to choose between the available datetimes (URI-R$\MVAt t_{1} \Rightarrow$ URI-T $\Rightarrow$ URI-R$\MVAt t_{2}$). 
\subsection{Pattern 3: Dive} 
Dive is when a user accesses different URI-Rs at nearly the same datetime. In this pattern, the user accesses one URI-R at a specific time, URI-R$_1 \MVAt t_0$, then navigates to different hyperlink(s) of URI-R$_{1}$'s page (e.g., URI-R$_{2} \MVAt t_{0}$) and so on (Figure \ref{fig:P2}).
\begin{eqnarray}
Dive &=&\{\mbox{URI-X}_i|\:i>1\mbox{, URI-X}\in \mbox{\{URI-T, URI-M}\} \nonumber \\
	  & & \mbox{and URI-R(URI-M}_{i}) <> \mbox{URI-R(URI-M}_{i-1})\} \nonumber
\end{eqnarray}
\subsection{Pattern 4: Skim} 
Skim is when a user accesses a number of different TimeMaps for different URI-Rs (Figure \ref{fig:P4}). Skim does not include any access for mementos. 
\begin{eqnarray}
Skim &=&\{\mbox{URI-X}_i|\: i > 1 \: \mbox{and URI-X} \in  \{\mbox{URI-T}\}\} \nonumber
\end{eqnarray}

\section{Methodology}
In this study, we introduce an analysis of the user access patterns of web archives. The analysis was conducted on the Internet Archive's Wayback Machine access logs. The first step in preparing the Wayback access logs for usage mining was transforming the raw log file into server sessions through web-log preprocessing, which included data cleaning, user identification, and session identification \cite{RobertCooley1999}. Then, we performed feature extraction, robot detection, and user access pattern detection.
\subsection{Wayback Machine Access Logs}
A Web server log file is a plain text file that records the activity information of the submitted requests from the users on the web server. The Wayback Machine access logs contain the following fields: client IP, access time, HTTP request method (GET or HEAD), URI, protocol (HTTP), HTTP status code (200, 404, etc.), bytes sent, referring URI, and User-Agent. For privacy purposes, the Internet Archive anonymized the client IP address. A segment from the Wayback Machine server logs, which we will call Wayback access logs, is shown in Figure \ref{fig:sample}. The first line shows a request for a URI-T. The second line shows a request for a URI-M.
\begin{table*}
  \centering
    \begin{tabular} {|l| r r r r r r r|r r r| }\hline 
    \textbf{Days} & \textbf{Feb 2} & \textbf{Feb 3} & \textbf{Feb 4} & \textbf{Feb 5} & \textbf{Feb 6} & \textbf{Feb 7} & \textbf{Feb 8} & \textbf{Mean} & \textbf{SD} & \textbf{SE} \\ \hline \hline
\textbf{Duration}&0:33:12&0:31:15&0:40:34& 0:42:57 & 0:29:35 & 0:25:45 & 0:24:33 & 0:32:33 & 0:06:29 & 0:02:27 \\ \hline
\textbf{GET}&98.4\% & 99.3\% & 97.7\% & 97.9\% & 99.4\% & 99.7\% & 99.8\% & 99\% & 0.8\% & 0.3\% \\ \hline
\textbf{Embedded}&47.4\% & 34.8\% & 43.7\% & 42.7\% & 41.9\% & 44.7\% & 46.8\% & 43.1\% & 3.9\% & 1.5\% \\ \hline
\textbf{SI Robots}& 6.2\% & 12.0\% & 7.7\% & 7.7\% & 2.9\% & 3.5\% & 3.8\% & 6.3\% & 3.0\% & 1.1\% \\ \hline
\textbf{NullRef}&42.6\% & 56.6\% & 47.5\% & 47.0\% & 49.4\% & 42.6\% & 43.9\% & 47.1\% & 4.6\% & 1.7\% \\ \hline
\textbf{s2xx}&33.7\% & 32.4\% & 34.2\% & 33.2\% & 34.1\% & 33.4\% & 33.6\% & 33.5\% & 0.6\% & 0.2\% \\ \hline
\textbf{s3xx}&51.8\% & 52.3\% & 50.8\% & 52.2\% & 51.7\% & 51.9\% & 53.2\% & 52.0\% & 0.7\% & 0.3\% \\ \hline
\textbf{s4xx}&11.7\% & 13.1\% & 12.0\% & 11.6\% & 11.2\% & 10.3\% & 10.1\% & 11.4\% & 0.9\% & 0.4\% \\ \hline
\textbf{s5xx}&2.8\% & 2.3\% & 3.0\% & 2.9\% & 3.0\% & 4.4\% & 3.1\% & 3.1\% & 0.6\% & 0.2\% \\ \hline
\textbf{Cleaned}&21.3\% & 23.0\% & 17.6\% & 17.7\% & 20.7\% & 18.1\% & 16.9\% & 19.3\% & 2.2\% & 0.8\% \\ \hline
\textbf{Sessions}&37,634 & 31,731 & 32,159 & 28,750 & 36,087 & 35,848 & 32,117 & 33,475 &  2,896 &  1,094 \\ \hline
   \end{tabular}
  \caption{Features for each sample of 2M records, Feb.\@ 2-8, 2012.}
  \label{tab:stat}
\end{table*}
\subsection{Dataset}
The Wayback Machine allows users to browse archived copies of web pages across time. The Wayback access logs were sampled using two probability techniques \cite{Teddlie2007}: cluster sampling, which is choosing a cluster of data randomly, and random sampling, where each sampling unit has an equal chance of being included. We performed cluster sampling by choosing a week (Feb.\@ 2-8, 2012) and random sampling by taking a random slice from each day of the week. Each sample comprised a slice of 2M requests to the Wayback Machine web server. Table 1 shows the characteristics of each dataset. It contains the following features for each sample:
\begin{itemize}
\item \textbf{Duration:} the difference between the last request time and the first request time of each sample in HH:MM:SS format.
\item \textbf{GET:} the percentage of requests that used the GET method.
\item \textbf{Embedded:} the percentage of requests that were for embedded resources of web pages (such as images and CSS files, etc.).
\item \textbf{SI Robots:} the percentage of requests by self-identified robots based on the User-Agent field.
\item \textbf{NullRef:} the percentage of requests that had an empty referral field.
\item \textbf{s2xx:} the percentage of successful requests (2xx status code).
\item \textbf{s3xx:} the percentage of redirections (3xx status code).
\item \textbf{s4xx:} the percentage of client errors (4xx status code).
\item \textbf{s5xx:}: the percentage of server errors (5xx status code).
\item \textbf{Cleaned:} the percentage of requests remaining after removing requests for embedded resources, HEAD requests, and requests that resulted in status codes other than 200, 404, or 503. 
\item \textbf{Sessions:} the number of sessions.
\end{itemize}

The last three columns of the table show the mean, standard deviation, and corresponding standard error between the samples. We use the Feb. 2, 2012 sample in our analysis because as we see from Table \ref{tab:stat}, it is a representative sample. 

In the Feb. 2 sample, we note that HTTP 3xx accounts for 52\% of the total number of requests. This is related to the default Wayback Machine behavior. First, the Wayback Machine rewrites all of the hyperlinks of a memento's embedded resources with the mementos's timestamp. Second, in the resolution of these URIs, the Wayback Machine will redirect the request of the embedded resources and hyperlinks to the nearest (timestamp) available memento. Furthermore, the Wayback Machine responds with a 302 status first when the requested URI-R is not available on the Wayback Machine, and then responds with a 404 status. 

\subsection{Data Cleaning} 
The first step in preprocessing our dataset was data cleaning, i.e., removing log entries that were not needed for the mining process \cite{Markov2007, Tanasa2004}. In similar studies for log analysis, robots that identify themselves in the User-Agent field were removed. Because robots crawl web archives intentionally, we did not exclude their requests in the cleaning step. We eliminated the following items which were irrelevant in terms of user behavior:
\squishlist
\item Requests that were generated automatically by the web browser for embedded resources of the requested web page (such as graphic files, page style files, etc.).
\item Entries with an HTTP status code other than HTTP 200, 404, or 503. We kept only these because we considered them to be requests executed by the user.
\item Requests using the HEAD request method (as suggested by \cite{Liu:2007:CMW:1231540.1231807}).
\item Static resources of the Internet Archive web site and the URIs of the liveweb service, which the Internet Archive introduces to redirect the user to the live web when the copy is not found on the Wayback Machine.
\item Invalid requests from web sites which included a link for malformed URI-Rs (for example, about:blank) among their embedded resources, so that each request on their web sites caused automatic requests to the Wayback Machine server. Similar behavior had been detected by Omodei \cite{Omodei2012}.
\squishend
\begin{figure}[H]
\centering
	\includegraphics[width=3in, height=2.7in]{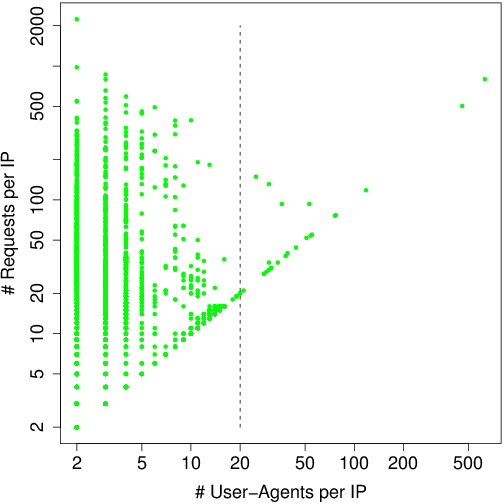}
	\caption{The number of user requests per IP against the number of User-Agents per IP.}
	\label{fig:UASL1}
\end{figure}
\subsection{User Identification}
To identify the users, the log files were first sorted by the IP, then by the request time. At first, we identified users by the 2-tuple (IP, User-Agent), but we found instances of malicious robots who not only did not self-identify, but who also changed their User-Agent with every request. There are some legitimate cases for humans to have different User-Agents and the same IP, for instance, two simultaneous users coming from behind a firewall. So, we needed to determine a reasonable threshold for the number of User-Agents per IP to detect malicious robots. 

Figure \ref{fig:UASL1} shows the relationship between the number of User-Agents per IP and the number of requests per IP (for those IPs with more than one User-Agent). We excluded self-identified robots from the graph to avoid biasing the results. We found only 24 users who had changed their User-Agent field more than 20 times. The median value for the number of User-Agents per IP for the dataset is 3 (excluding users who had one User-Agent only). We concluded that 20 different User-Agents per IP is a good threshold for this dataset. For IPs with at most 20 different User-Agents, we used the 2-tuple (IP, User-Agent) to identify individual users. The 24 IPs with more than 20 different User-Agents were classified as 24 separate robots.

\begin{table*}
\centering
\begin{tabular}{lrr}
\hline
&\multicolumn{2}{c}{\textbf{\# Detected Robots}} \\
\cline{2-3}
\textbf{Heuristics} & \textbf{out of 37,634 sessions} & \textbf{out of 426,317 requests} \\ \hline
\textbf{SI Robots} & 1,410 & 68,967 \\
\textbf{\#UA per IP }& 24 & 2,747 \\
\textbf{Robots.txt} & 55 & 90 \\
\textbf{Browsing Speed} & 1,601 & 47,320 \\
\textbf{Image-to-HTML} & 33,244 & 326,019 \\  \hline
\textbf{Total Robots} & \textbf{34,203 (90.9\%)} & \textbf{396,627 (93.9\%)} \\
\hline
\end{tabular}
\caption{The number of detected robots from applying each heuristic independently and the number of the records after applying all the filters together.}
\label{tab:filters2}
\end{table*}
\subsection{Session Identification} 
A session is the set of web pages that are requested by particular user \cite{Markov2007}. Session identification is performed by dividing a web server log file into web server sessions. First, we group all the requests based on the IP and User-Agent (as described in Section 4.4). Second, we apply a threshold timeout, so that if the time elapsed between two consecutive requests is longer than this threshold, the second request is considered to be the first request of the new session. There have been several suggested timeout thresholds including 25.5 minutes \cite{Catledge1995}, 30 minutes \cite{Tan2002, Kumar2010}, and 60 minutes \cite{Anick:2003:UTF:860435.860453}. Others proposed 10 minutes as a conservative threshold to capture the time for staying on one page \cite{Liu:2007:CMW:1231540.1231807, Spiliopoulou2003}. In our study, we divided the requests of each user into individual sessions based on a 10 minute timeout threshold. Future research is required to verify that searching and browsing models for the current web are valid for browsing the past web (i.e., web archives).

After identifying the sessions, we extracted features for each session to be used further in the analysis. A session, S, is 7-tuple. 
\begin{eqnarray}
S = \langle URI, S_l, S_d, BS,\bar{S}_{rt} , stdev(S_{rt}), IH \rangle \nonumber  
\end{eqnarray}
The following is the description of each item:
\squishlist 
\item \textit{URI} is the set of URIs that the user visited in the session. The set of URIs are defined as:
\begin{eqnarray}
URI &=&  \{ \mbox{URI}_i|\mbox{ } i \mbox{ is an integer, }1 \leq i\leq S_l  \nonumber \\
    & & \mbox{and } \mbox{URI} \in \mbox{\{URI-T, URI-M\}}\} \nonumber  
\end{eqnarray}
\item $S_{l}$, session length, is the number of webpages the user requested during the session.
\item $S_{d}$, session duration, is calculated by subtracting the timestamp of the first request of the session from the timestamp of the last request of the session.
\item \textit{BS} is the browsing speed of each session in requests/second. $BS = S_{l} / S_{d}$.
\item $\bar{S}_{rt}$ is the mean inter-request time of the session. 
\item $stdev(S_{rt})$ is the standard deviation of the inter-request time of the session.
\item \textit{IH}, image-to-HTML, is the ratio between the number of image files and the number of HTML files per session.
\squishend
\begin{table}[h!]
\centering
\begin{tabular}{|l|>{\raggedleft\arraybackslash}p{3.7cm}|}
\hline
\textbf{Filters} & \textbf{\% Excluded Requests (out of 2M)} \\ \hline
Status Code & 51.8\% \\
Embedded Resources & 47.4\% \\
Static and Liveweb & 10.0\% \\
Invalid Requests & 3.7\% \\
HEAD & 1.6\% \\
 \hline
\textbf{All Filters} & \textbf{78.7\%} \\
\hline
\end{tabular}
\caption{The characteristics of data cleaning filters. Some of requests fell into multiple categories, so the percentages add up to more than 100\%.}
\label{tab:filters1}
\end{table}

\subsection{Robot Detection}
Because of the increasing numbers of web crawlers that are engaged in web harvesting, many studies have been conducted for investigating the robot detection problem \cite{Tan2002, Kwon2012}. In this study, we used different types of robot detection techniques \cite{Doran2010}. First, we applied syntactical log analysis by checking the User-Agent field to identify the self-identified robots. Second, we applied traffic pattern analysis techniques to distinguish humans from robots based on their navigational behavior. In this section, we describe heuristics we used for distinguishing robots from humans.

\subsubsection{User-Agent Check} 
The User-Agent check is applied for requests from crawlers and robots which declared their identity to the web server through the User-Agent field (SI robots). We excluded these robots by applying this heuristic before the calculations of the session features to avoid biasing the results.

\subsubsection{Number of User-Agent per IP}
As explained earlier (Section 4.4), we used 20 as a threshold for the maximum number of different User-Agents for each IP. Users with more than 20 different User-Agents were classified as robots.

\subsubsection{Robots.txt file}
Web site administrators put a list of access restrictions to specify which parts of their web site are not allowed to be visited by robots. We labeled the sessions in which users downloaded the robots.txt file for the Wayback Machine (http://web.archive.org/robots.txt) as robots. 
\subsubsection{Browsing Speed (BS)} 
The importance and the effect of \textit{BS} has been discussed and used for detecting robots many times \cite{Nithya2012, Tanasa2004, Reddy2012}. We use $BS \leq 0.5$ (i.e., no faster than one request every two seconds) as a threshold for human browsing speed \cite{Castellano2007}. We observed that this threshold is appropriate for our dataset, so we classify the sessions with $BS > 0.5$ as robots.

\subsubsection{Image-to-HTML Ratio (IH)} 
Human sessions should have more images than robot sessions because of the embedded images present in most HTML pages. Robots tend to retrieve only HTML pages, while ignoring image formats. We used the $IH$ metric calculated previously for each session to detect robots. In \cite{Stassopoulou2009}, 1:10 $IH$ had been suggested as a good threshold for distinguishing robots from humans. We label a session with less than one image file for every 10 HTML files as a robot. $IH$ is the only heuristic that does not require a session have at least two requests. This heuristic is the best predictor for robots \cite{Tan2002, Stassopoulou2009}, and it has a strong effect on our dataset.

\section{Results and Analysis}
In this section, we explain the results of preprocessing the dataset (described in Sections 4.3-4.5) and of applying the heuristics for robot detection (described in Section 4.6). We analyze the resulting data and contrast the behavior and access patterns of humans and robots. We conclude with an analysis of the temporal preference of human users.

\begin{table*}[ht!]
\centering
\begin{tabular}{|l| >{\raggedleft\arraybackslash}p{0.9in} |>{\raggedleft\arraybackslash}p{1in} |>{\raggedleft\arraybackslash} p{0.9in}| >{\raggedleft\arraybackslash}p{0.95in} |r| r|} 
\hline
\textbf{Users} & \textbf{\# Requests (Filtered)} & \textbf{\# Requests (Raw)} & \textbf{\# Sessions}  & \textbf{\# Transferred MB} & \textbf{\# URI-Ts} & \textbf{\# URI-Ms}\\ \hline
\textbf{Robots}  & 396,627 (93.0\%) & 1,002,573 (50.1\%)& 34,203 (90.9\%) & 20,010 & 378,201 (95.4\%) &  18,426 (4.6\%) \\
\textbf{Humans}  & 29,690 (7.0\%) & 810,049 (40.5\%)& 3,431 (9.1\%) & 4,459 & 5,505 (18.5\%) & 24,185 (81.5\%)\\
\hline
\end{tabular}
\caption{HTTP activity of robots and humans.}
\label{tab:robhu}
\end{table*}

\subsection{Traffic Analysis}
To extract the user access patterns for web archives from the Wayback access logs, we first applied data preprocessing techniques (data cleaning, user identification, session identification) to convert the log file into web server sessions. The raw log file contains 2M requests from which we determined 21,932 unique IPs. Because of the stateless nature of the log files, we identified the users based on the IP and User-Agent to identify 33,841 users who created 37,634 different sessions.
 
The characteristics of each filter (3xx status code, embedded resources, static resources and liveweb, invalid requests, HEAD) and the total number of excluded requests after applying all the filters together are shown in Table \ref{tab:filters1}. The number of records in the Feb. 2 sample was decreased from 2M to 426,317 (21.3\% of the requests in the raw file).
\begin{figure}[h!]
	\includegraphics[width=3.3in]{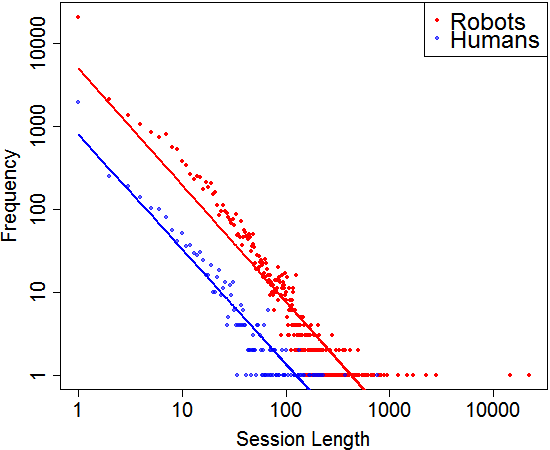}
	\caption{The frequency of session lengths (\# of requests) for humans and robots.}
	\label{fig:SL}
\end{figure}
\subsection{Robots vs Humans}
Table \ref{tab:filters2} contains the results of applying the heuristics for detecting robots. The rules are not mutually exclusive, but we calculated the number of requests which had been labeled as robots from each filter separately. $IH$ had the largest effect on detecting robots. We used a 1:10 $IH$ as a threshold for distinguishing robots from humans. We found that 99.93\% of the sessions which were detected by this heuristic had 0 images. \textit{BS} is also important, because it classified a significant number of robots who had a \textit{BS} (more than 0.5 requests/second) impossible for humans.

Table \ref{tab:robhu} contains the summary of the activity of humans and robots. From the table, we notice that the sum of the percentage of raw requests from humans and robots did not equal 2M requests. The reason is that there are many accesses that were created by invalid requests to the web server. Furthermore, there are many requests to embedded resources only, which were filtered. The percentage of human requests after cleaning and separating robot requests is only 1.5\% of the 2M requests.

The significant discovery here is the 10:1 ratio of robot sessions to human sessions. This ratio is a strong motivation for building an API interface that serves robot accesses in order to decrease the load of robots on the Wayback Machine. A typical human session costs more than a robot session as humans average 1.30 MB/session and robots average 0.58 MB/session. Human requests include automatic downloads of the embedded resources of the web pages they access, and robots usually ignore downloading these embedded resources.

We discovered that most of the robots had a breadth search strategy in downloading the web pages from the Wayback Machine; more than 95\% of the robots downloaded TimeMaps only, as shown in Table \ref{tab:robhu}.
On the other hand, of all human requests, only 18.5\% were for TimeMaps. 

\begin{figure}[ht!]
	\includegraphics[width=1\linewidth, height = 3in]{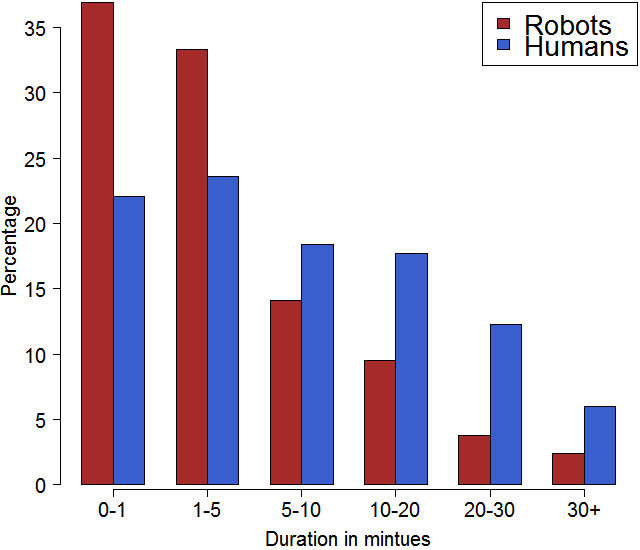}
	\caption{The percentage of sessions for each interval. The number of humans and robots sessions are 2024 and 17019 (excluding the sessions with one request), respectively.}
	\label{fig:sd}
\end{figure}

\subsubsection{Session Length $(S_{l})$}
After detecting the robots, we separated them from humans and analyzed their behaviors individually. Figure \ref{fig:SL} shows the session length frequency for robots in red and for humans in blue. We notice from the figure that many more robots have longer sessions than humans. The $\bar{S}_{l}$ for robots is 10 requests/session, while humans have an $\bar{S}_{l}$ of 9 requests/session.

\subsubsection{Session Duration $(S_{d})$}
We computed session duration by subtracting the time of the first request from the time of the last request for each session. Session duration requires at least two requests. Figure \ref{fig:sd} shows the percentage of sessions with different session durations for robots and humans. We can see that the majority of the sessions were short, taking into consideration that we did not count the time spent on the last requested web page. The $\bar{S}_{d}$ is 10 minutes for robots and 5 minutes for humans. 

\subsubsection{Inter-Request Time}
We also calculated $\bar{S}_{rt}$ and $stdev(S_{rt})$ for each session. We found that the median values of $\bar{S}_{rt}$ for human and robot sessions are 19 seconds and 40 seconds, respectively. The median of $stdev(S_{rt})$ values is 37 seconds for robots and 11 seconds for humans. This indicates that robots tend to have more irregular periods between HTML requests than humans, and this matches the finding by Tan et al.\@ \cite{Tan2002}.

\begin{figure*}
	\centering 
	\subfigure[Robots (34203 sessions)]{
	\includegraphics[width=0.4\linewidth]{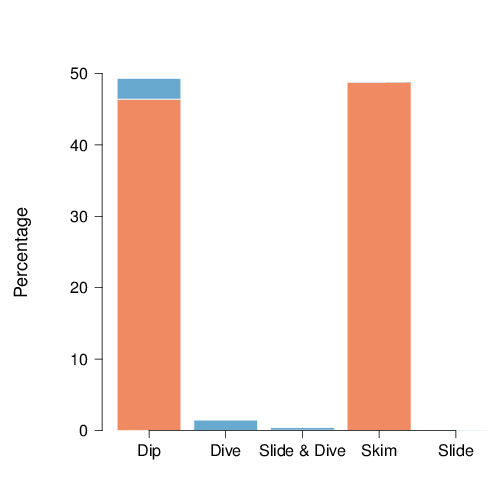}
	\label{fig:pcr}
	}
	\subfigure[Humans (3431 sessions)]{
	\includegraphics[width =0.4\linewidth]{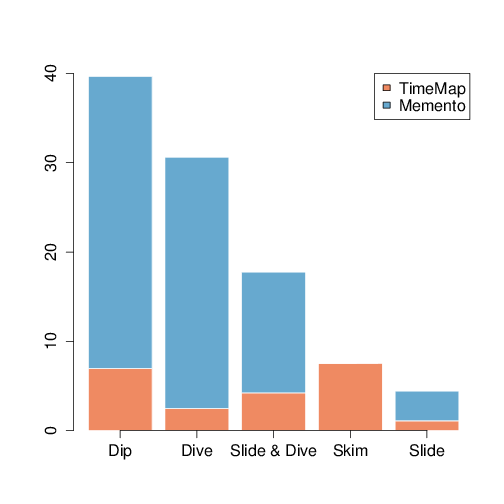}
	\label{fig:pch}
	}
	\caption{Robots and humans exhibit different access patterns.}
	\label{pc}
\end{figure*} 
\subsection{Web Archive User Access Patterns}
How do users go through web archives? Do they go in deeply from URI-R$_{1}$ to URI-R$_{2}$, do they browse broadly from URI-M$_{1}$ to URI-M$_{2}$ for the same URI-R, or do they use a combination of these two patterns? Are robot accesses similar to human accesses?

In this section, we answer the previous questions by extracting the user access patterns for web archives from our filtered dataset. The requested URIs for each session were extracted and then identified based on their type, URI-M or URI-T. We also extracted the URI-R of each requested URI to compare it with the other URI-Rs from the same session. Because of the existence of different forms of URIs which refer to the same website \cite{McCown2006}, we applied URI canonicalization for the URI-Rs to normalize them under one host \cite{Cutts2006}.

We discovered four basic building blocks (Dip, Slide, Dive, Skim) of the user access patterns for web archives. Figures \ref{fig:P0}-\ref{fig:P4} show the models along with examples of the four patterns. The percentages of each pattern exhibited in robot and human sessions are shown in Figures \ref{fig:pcr} and \ref{fig:pch} along with the percentages of requests to TimeMaps and mementos for each pattern. 

\subsubsection*{Dip} Dip is the pattern where the user requests a single URI. This URI can be a URI-M or a URI-T. It represents the most repeated pattern for humans (33\% of all sessions) and robots (49\% of all sessions). URI-Ms contribute to 83\% of human sessions that exhibit the Dip pattern, although 94\% of the robot Dips are requests for URI-Ts.

\subsubsection*{Slide} The user who is interested in travelling through time, browsing different copies of the same URI-R, creates the Slide pattern. There are only a few humans who access the web archives broadly then navigate away (4.2\% of all sessions). Robot sessions do not have this pattern with a noticeable percentage (0.1\% of all sessions).

\subsubsection*{Dive}  
Dive represents the second highest percentage of human sessions, 29.7\%. In this pattern, the user goes deeply for browsing hyperlinks of URI-Ms. The robot sessions which were composed of this pattern crawl the web sites deeply, but they are not a significant number of sessions. 

\subsubsection*{Skim} Skim is the pattern for which the users access different numbers of TimeMaps. Robot sessions exhibit this pattern 48.7\% of the time. Investigating the relationship between the topics of the URI-Rs of the requested TimeMaps during a single session is one of our goals for upcoming research.
\begin{table}[H]
\centering
\begin{tabular}{|l|l |rrr|} 
\hline
\textbf{User} & \textbf{Pattern} & \textbf{Median} & \textbf{Mean} & \textbf{SD} \\ \hline
 & Slide & 3 & 3 & 1.4 \\
\textbf{Robots} & Dive & 3 & 15 & 53.2 \\
 & Skim & 3 & 21 & 267.0 \\ \hline
 & Slide & 3 & 4 & 3.4 \\
\textbf{Humans} & Dive & 4 & 8 & 14.3 \\
 & Skim & 3 & 6 & 7.2 \\
\hline
\end{tabular}
\caption{Statistics for the length of all Slides, Dives, and Skims}
\label{tab:pstat}
\end{table}

\subsubsection*{Slide and Dive}
A large number of human sessions consist of at least two occurrences of the Dive and Slide patterns. In these sessions, the users request URI-R$_{1}$ and browse its different copies at different times (URI-R$_{1} \MVAt t_{1}$ $\Rightarrow$ URI-R$_{1} \MVAt t_{2} \Rightarrow$ URI-R$_{1} \MVAt t_{3}$), then dive through a hyperlink (URI-R$_{2} \MVAt t_{3}$) from URI-R$_{1} \MVAt t_{3}$, then repeat Dive or Slide. In contrast, users may start by going deeply through different mementos for different URI-Rs (Dive pattern), then go broadly through one of these mementos to browse other captures at different times (Slide pattern) (e.g., URI-R$_{1} \MVAt t_{1}$ $\Rightarrow$ URI-R$_{2} \MVAt t_{1} \Rightarrow$ URI-R$_{3} \MVAt t_{1} \Rightarrow$ URI-R$_{3} \MVAt t_{2}$, etc.). The percentage of human sessions that were composed of a combination of these two patterns is 17.2\%. We calculated the number of Slides and Dives for these sessions and found 1167 Slides and 1942 Dives. For robot sessions that were composed of Slide and Dive, we found 328 Slides and 571 Dives.
\begin{figure}
\centering 
	\includegraphics[width=3.1in, height=2.5in]{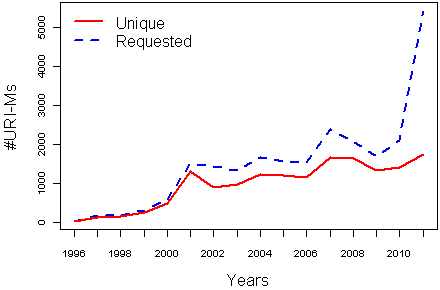}
		\caption{Distributions of the years for the unique and requested mementos by humans.}
	\label{fig:yearsur}
\end{figure}		

\subsubsection*{Pattern Length}
Each pattern is made up of a number of requests, which we call the pattern length. We calculated the pattern length for all sessions. The median, mean, and standard deviation of the lengths of each pattern for robots and humans are summarized in Table \ref{tab:pstat}. For humans, the longest mean pattern length is an 8-request Dive, while for robots the longest mean pattern length is 21 requests in a Skim.

\subsection{Temporal Analysis}
Figure \ref{fig:yearsur} shows both the unique and total number of mementos referenced grouped by the year of their Memento-Datetime. Although there is no clear temporal preference for any one year of the unique mementos, there were a significant number of repeated requests for mementos from 2011. This locality of reference suggests that there is an important benefit to be gained by caching the mementos from the recent past. Figure \ref{fig:yearsua} shows that the total number of mementos available for 2011 was similar to previous years. In both Figures \ref{fig:yearsua} and \ref{fig:yearsur}, pre-2001 data is included although in those years the archives are too sparse for meaningful comparison with later years.

\section{Conclusions and Future Work}
We introduced the basic building blocks (Dip, Slide, Dive, and Skim) for user access patterns for web archives through an analysis of the Internet Archive's Wayback Machine access logs. We applied heuristics for detecting robots and found that robot sessions outnumber human sessions 10:1. This suggests that there is utility in building an API interface that serves robot accesses. Robots account for 91\% of sessions and 93\% of requests to the Wayback Machine, yet robots outnumber humans 5:4 only in terms of raw, unfiltered requests and 4:1 in terms of megabytes transferred. We found that humans download more information per session due to embedded resources, which robots ignore. We also analyzed human and robot access patterns to emphasize the similarities and the differences between them. We found that robots mainly exhibit the Dip and Skim patterns, with about 49\% of their sessions for each pattern, and that they access TimeMaps almost exclusively. Humans exhibit the Dip and Dive patterns the most with 39\% and 30\% of their sessions, respectively. Unlike robots, humans mainly access archived pages rather than TimeMaps. Finally, we provide an analysis for the temporal preferences of humans based on the Memento-Datetime (by year) of their requests and discovered significant repetitions for requests in 2011. This suggests that there is a benefit to be gained by caching mementos from the recent past.

Web server logs are a rich source for information about web archives. We are planning to extend our analysis to serve other applications of web usage mining, such as personalization for making dynamic recommendations to web archive users based on their navigational behavior patterns by using data mining techniques. Further, we will study the validity of applying searching and browsing models for the current web to web archives. We also expect that Slides and Dives that users create on web archives may create stories around a particular event. We plan to extend our study on a large data set to detect stories that humans might create from their access patterns, which will be integrated into the live web to produce automatic stories about specific event for the users.
\begin{figure}
\centering 
	\includegraphics[width=0.65\linewidth, angle=270]{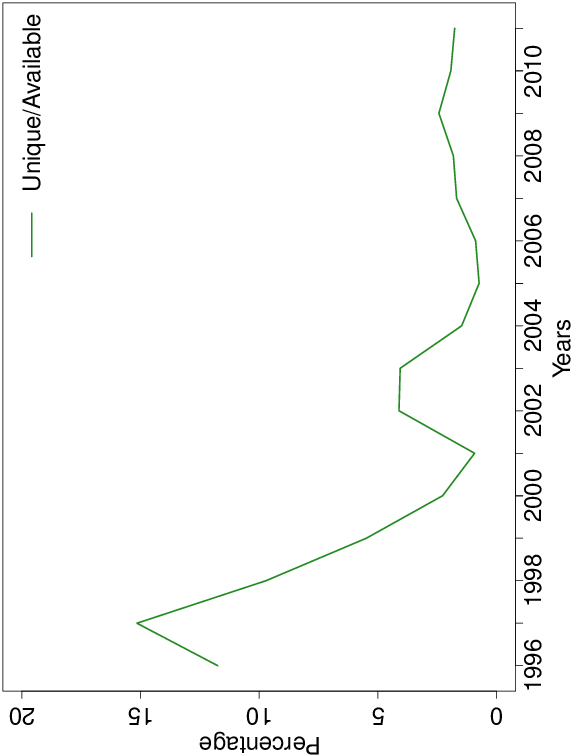}
	\caption{The proportion of unique URI-Ms requested out of the potential requested for each year.}
	\label{fig:yearsua}
\end{figure}
\section{Acknowledgments}
This work was supported in part by the NSF (IIS 1009392) and the Library of Congress. We thank Kris Carpenter Negulescu (Internet Archive) for access to the anonymized Wayback Machine logs.
 
%
%
\bibliographystyle{abbrv}
\bibliography{fp105-AlNoamany} 
\balancecolumns
\end{document}